# Single-molecule electron diffraction imaging with charge replacement

E.E. Fill[1], F. Krausz[1,2] and M. G. Raizen[3]

[1]*Max-Planck-Institut für Quantenoptik, Garching, D-85748 Germany*

[2]*Ludwig-Maximilians-Universität München, Garching, D-85748 Germany*

[3]*Center for Nonlinear Dynamics and Dept. of Physics, The University of Texas at Austin, Austin, Texas 78712, USA*

**Abstract.** We investigate the possibility of non-destructive electron diffraction imaging of a single molecule to determine its structure. The molecular specimen will be held on a free-standing sheet of graphene. Due to the high conductivity of graphene, electrons lost by ionization would be rapidly replaced, enabling repeated nondestructive interrogation. Limits of resolution, maximum particle size and required electron flux are assessed.

# 1. Introduction

The determination of protein structure is one of the grand challenges of structural chemistry and biology. The standard methods of interrogation to date are X-ray and electron diffraction, but the high energies required to yield sufficient resolution are destructive at the level of individual proteins and therefore require ensemble averaging on aligned molecules. Therefore, successful determination of atomic structure has been limited so far to a relatively small set of proteins that can be crystallized [1].

It is well-known that the basic limitation in the determination of molecular structure by imaging is radiation damage [2-5]. For both, X-rays and electrons inelastic cross-sections are significantly higher than elastic ones. For electrons the situation is more favourable since their ratio is only a factor of two to three [4]. Yet, the maximum allowed exposure of 500 electrons/nm$^2$ (for an electron energy of 100 keV) [3] is still some three orders of magnitude lower than that required by the Rose criterion for a carbon atom to be detectable against a background of shot noise.

To overcome this problem, two different approaches have been proposed to determine the three-dimensional structure of bio-molecules without the need for crystallizing them. These are X-ray diffraction of single molecules with ultrashort pulses from a free-electron X-ray laser (FEL) [6] and continuous-wave (cw) electron diffraction of molecules aligned by a laser [7]. In the former, a diffractogram of a molecule would be recorded in a single shot, taking advantage of the high number of photons in the X-ray laser pulse. Although the molecule is destroyed with each shot, the short pulse duration of a few femtoseconds (fs) could generate an image before the molecule disintegrates. A special algorithm [8] ensures that only molecules with accidental alignment are recorded. A 3D image is generated by evaluating pictures taken at different angles of alignment. In the second proposal, the molecules would be frozen in a helium [7] or water [9] droplet and delivered to the interaction chamber at a high rate. In both proposals diffractograms of aligned molecules would be obtained; phase retrieval algorithms could then evaluate the diffraction patterns [10, 11]. Both methods result in destruction of the molecule by ionization and subsequent Coulomb explosion [6]. Furthermore, it is a formidable challenge to generate few-fs intense X-ray pulses or laser-align molecules encapsulated in frozen droplets.

In this paper we propose that the atomic structure of molecules could possibly be probed non-destructively by attaching the molecule to a sheet of single-layer graphene. Graphene consists of carbon atoms densely packed in a honeycomb crystal lattice [12-14]. It has remarkably high stability and, moreover, a freestanding sheet of only one atomic layer can be suspended over a mesh containing holes of microns in diameter [15]. Graphene is almost invisible for electrons and has a well-defined diffraction pattern [13]. Transmission electron microscope (TEM) imaging of single molecules on graphene membranes has already been demonstrated [16]. Most importantly for our approach, its conduction band electrons exhibit an extraordinarily high mobility, propagating ballistically over distances of several hundred nanometers [12].

In what follows: we derive parameters of the electron source and consider limits of resolution and sample size. We then analyze the issue of charge replacement in the molecules being are ionized, which is crucial for avoiding Coulomb explosion or breaking of bonds over the extended periods of exposure.

## 2. Resolution, probe size and brightness

The basic arrangement for diffractive imaging of a molecule on a grapheme sheet is shown in fig. 1. The molecule is irradiated with a cw electron beam. The direct beam is blocked in front of the detector while elastically scattered electrons are recorded and generate a diffraction pattern. The pattern is evaluated by means of a phase retrieval algorithm to find a projection of the molecule [10, 11]. The 3D structure of the molecule is then found by evaluating a sufficient number of such projections around a tilt axis [17].

In diffractive imaging the maximum scattering angle $\Theta_{max}$ from the sample to the detector determines the maximum observable momentum transfer wave number $q_{max}$ [18, 19]. Defining the resolution length as $r_l = 1/q_{max}$, and using the fact that $q = \frac{2}{\lambda_{dB}} \sin \Theta/2$ for elastic scattering, we obtain

$$r_\ell = \lambda_{dB} / \Theta_{max}, \tag{1}$$

where $\lambda_{dB}$ stands for the de Broglie wavelength of the electrons.

One further needs to consider the maximum sample size $w_{max}$, which is limited by spatial frequency sampling as well as by longitudinal and transverse coherence. Consider Nyquist sampling in the Fourier plane: requiring that the highest spatial frequency on the detector be sampled by at least two pixels per period, i.e. $w_{max}/\lambda_{dB} z \leq 1/2\Delta x$, where $\Delta x$ is the width of a pixel and $z$ is the distance from the sample to the detector, one obtains the condition [18]

$$w_{max} \leq \lambda_{dB} z / 2\Delta x. \tag{2}$$

Longitudinal and transverse coherence are important parameters for diffractive imaging, and the longitudinal coherence length $l_c$ must be greater than the maximum path difference across the sample. Thus we require that $l_c > w\Theta_{max}$, where $w$ is the transverse width of the sample. For a particle beam, $l_c$ is determined by its energy spread: $l_c = \lambda_{dB}^2/\Delta\lambda = 2\lambda E/\Delta E$, where $\Delta\lambda$ and $\Delta E$ are the wavelength spread and energy spread, respectively, and $E$ is the energy of the particle (see, for example [20]). From this relation we derive for the relative energy spread

$$\Delta E / E = 2\lambda_{dB} / l_c < 2\lambda_{dB} / w\Theta_{max}, \tag{3}$$

and using equation (1) we obtain a relation containing only the resolution length and the particle size

$$\Delta E / E < 2\, r_l / w. \tag{4}$$

The particle size also yields a condition for spatial (transverse) coherence; the spatial coherence length of a particle beam is given by $l_s = \lambda_{db}/\varphi$, where $\varphi$ is the full angular spread of the particles [20]. The transverse coherence length must be larger than the sample width and thus

$$\varphi < \lambda_{db}/w. \tag{5}$$

To calculate the required flux of electrons and the dose on the sample we determine the probability of scattering of electrons into a pixel. For elastic scattering we get

$$P_{el} = N_a \frac{\partial \sigma}{\partial \Omega} \Delta \Omega, \tag{6}$$

where $N_a$ is the area density of scattering atoms, $\frac{\partial \sigma}{\partial \Omega}$ is the differential cross-section for elastic scattering and $\Delta \Omega$ is the spatial angle subtended by a pixel. To get a well-illuminated diffractogram we require a flux of 10 electrons per pixel even at the largest scattering angle determining the resolution. Thus we have for the number of electrons on the sample

$$N_e = 10 / \left[ N_a \left( \frac{\partial \sigma}{\partial \Omega} \right)_{\Theta_{max}} \Delta \Omega \right] \tag{7}$$

and, using $\Delta \Omega = (\Delta x / z)^2 = (\lambda_{dB} / 2 w_{max})^2$ (from eq. 2)

$$N_e = 40\, w_{max}^2 / \left[ N_a \left( \frac{\partial \sigma}{\partial \Omega} \right)_{\Theta_{max}} \lambda_{dB}^2 \right]. \tag{8}$$

This number of electrons must be supplied by an electron source with brightness $B$ to an area $w_{max}^2$, with a spatial angle $<(\lambda_{dB}/w_{max})^2$ (from eq. 5) and during a time $t_{acq}$, resulting in a required brightness given by

$$B > 40\, e\, w_{max}^2 \left/ \left[ N_a \left(\frac{\partial \sigma}{\partial \Omega}\right)_{\Theta_{max}} \lambda_{dB}^4\, t_{acq} \right] \right. \qquad (9)$$

As an example we consider the experimental requirements for a beam of electrons with an energy of 60 keV. The de Broglie wavelength at this energy is 0.005 nm. In a typical arrangement the detector is 10 cm from the sample and has a diameter of one inch, resulting in $\Theta_{max} \approx 100$ mrad. Equation (1) then yields a resolution length of 0.05 nm which is adequate for atomic resolution imaging. With a pixel width $\Delta x = 25$ μm the maximum sample size determined from equation (2) is $w_{max} = 10$ nm. Using this value, equations (4) and (5) yield $\Delta E/E < 10^{-2}$ and $\varphi < 0.5$ mrad. For a molecule with a width of 5 nm the corresponding values are $\Delta E/E < 2 \times 10^{-2}$ and $\varphi < 1$ mrad.

At 100 mrad the differential elastic scattering cross-section for 60 keV electrons on carbon is $1.58 \times 10^{-18}$ cm$^2$ / sr [21]. For a protein molecule we put $N_a = d\, N_c$ where $d$ and $N_c$ are the depth of the molecule and the number density of carbon equivalents respectively. The number density of carbon atom equivalents for a biomolecule is[4] $N_c = 6.6 \times 10^{22}$ cm$^{-3}$. For a spherical molecule $d = w_{max}$ and from eq. (8) the number of electrons on the sample becomes $N_e = 1.53 \times 10^9$. For the brightness required at an acquisition time of 10 s one obtains $B > 9.8 \times 10^7$ A/cm$^2$ s.

In fig. 2 brightness requirements and energy spread are displayed as a function of maximum molecule width $w_{max}$. The figure shows the case of a spherical molecule ($w_{max} = d$) and two cases in which the molecule is oblate ($w_{max} > d$) in the main part of the diagram. The required energy spread is shown for the case $w = w_{max}$. The requirements following from the above derivations are met by standard electron guns used in scanning electron microscopy (SEM) [22]. Field emission guns reach a brightness of $2 \times 10^8$ A/cm$^2$ sr with an energy spread of 0.2 - 0.4 eV. The small emission area allows focusing the beam down to a few nm using a single electron lens.

## 3. Preventing radiation damage

Inelastic scattering events can cause ionization of the target molecule. We consider under what circumstances a return flux of electrons from the graphene substrate could possibly prevent such damage. The number of ionization events is given by $N_{ion} = \sigma_{ion} N_a N_e$, where $\sigma_{ion}$ is the cross-section for ionization, and $N_e$ is the number of electrons needed for recording the diffractogram, as calculated above. The ionization cross-section of $CH_2$ (representing a biological molecule) for 60 keV electrons is [23] 2.1 x $10^{-18}$ cm$^2$. Inserting the values for $N_a$ and $N_e$ we arrive for a spherical molecule at a total of 2.1 x $10^8$ electrons which have to be replaced.

Such charge migration requires intramolecular charge transfer and charge transfer from substrate to molecule. The conductivity of biomolecules has been measured by several groups, but the reported characteristics vary over a wide range. Recently it has been shown that reproducible electronic coupling between the molecule and the probing electrodes is essential. Under this condition a conductivity of a single DNA molecule of $g = 0.1$ μS per base pair was measured [24]. Under normal conditions an adsorbed molecule is only weakly bound by van der Waals forces, in which case the electrons have to tunnel to the molecule. However, it has recently been shown that a biomolecule can be covalently bound to a carbon nanotube and that good electronic coupling can be achieved [25].

Assuming, therefore, that the bottleneck is the conductivity of the molecule, we can calculate the potential drop across a molecule subjected to a specified flux of electrons. The molecule is considered to be a thin nanowire, with contacts to the graphene substrate at both ends. If $N_{ion}$ electrons have to be replaced during a time $t_{acq}$, a current of $I_{mol} = N_{ion} e / t_{acq}$ is flowing through the molecule, where $e$ is the elementary electric charge. The potential across a base pair is then given by $U_{mol} = \alpha I_{mol} / g$. $\alpha$ is a prefactor of order 0.5 which takes into account that charge replacement occurs from both ends of the molecule.

Inserting the above numbers and assuming an acquisition time $t_{acq} \approx 10$ s we obtain a current of 3.4 pA through the molecule, resulting in a potential drop of about 17 µV per base pair. The resulting electric field is far below what could possibly destroy a molecule. Taking the length of a base pair to be 2.5 Å the resulting field strength is 7 µV/Å, which is almost a factor of $10^7$ below an atomic unit of field.

As already mentioned, if the molecule is not bonded to the substrate, the electrons have to tunnel from the substrate to the molecule. Tunneling times of electrons across van der Waals gaps have been determined [26]. For a vacuum gap a typical tunneling rate is found to be $10^{10}$ s$^{-1}$ at a distance of 5 Å and much higher at smaller distances. Using (pessimistically) the rate for a vacuum gap of 5 Å it turns out that charge replacement of 2.1 x $10^7$ electrons per second from the substrate would again be no problem.

## 4. Conclusion

Placing a molecule on an ultrathin substrate with high conductivity (such as graphen) may open the way for generating diffractograms of single biological molecules without their destruction. Our study reveals that electrons removed by the probing electron beam can be replaced by the substrate at a high enough rate to allow the molecule to survive the exposure required for a well-illuminated diffractogram. Verification of the method could be immediately tackled with state-of-the-art electron guns, which produce beams bright enough to record a projection of a 5 nm diameter molecule with 0.05 nm resolution in about ten seconds.

**Acknowledgement**

We thank Dr. W. Fuß for fruitful discussions. MGR acknowledges supports from the R. A. Welch Foundation and the Sid W. Richardson Foundation. He also thanks J. Hanssen, J. McClelland, E.-L. Florin, and P. McEuen for helpful discussions. This work was funded in part by the DFG Cluster of Excellence "Munich Centre for Advanced Photonics"-MAP (www.munich.photonics.de).

**Figures**

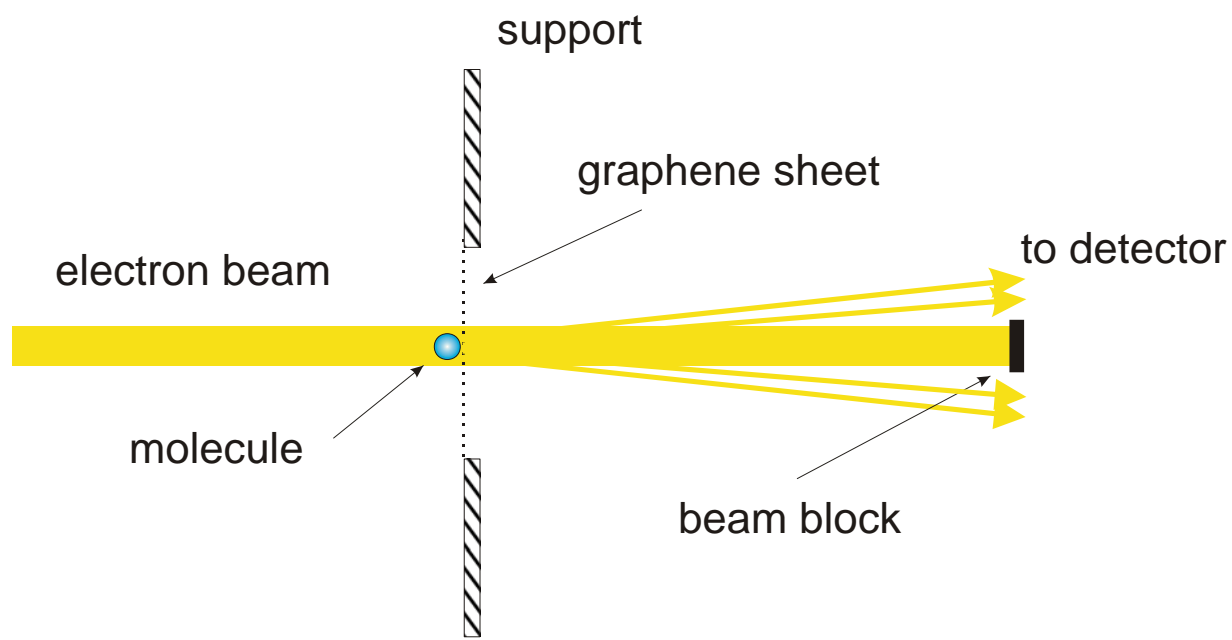

Fig. 1: Conceptual drawing of single molecule diffraction arrangement. An electron beam with a transverse coherence length exceeding the width of the molecule irradiates the molecule attached to a graphene sheet. The direct beam is blocked before the detector which records the diffraction pattern.

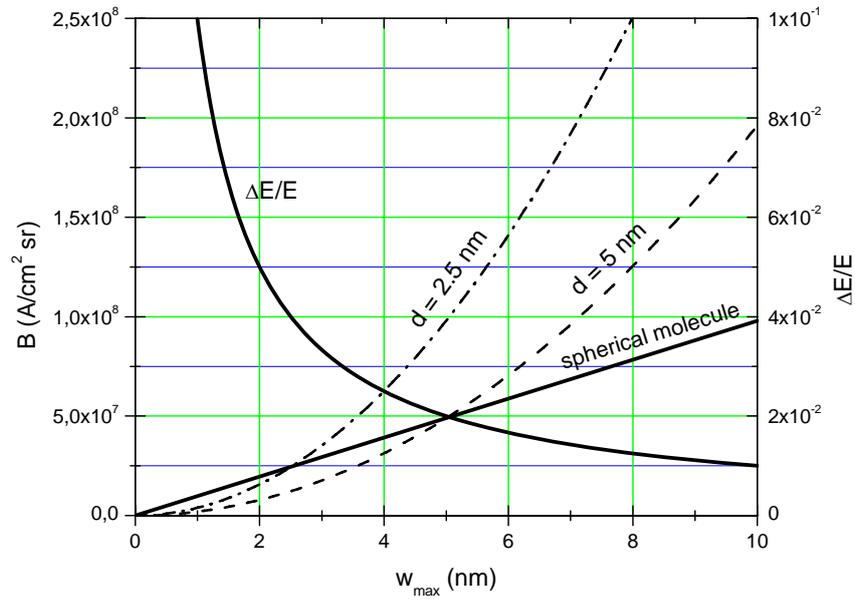

Fig. 2: Required brightness and allowed relative energy spread for diffraction imaging of a single molecule. Brightness is shown for a spherical molecule and for molecules with widths of 2.5 and 5 nm. Energy spread is the same for all molecular shapes. De Broglie wavelength 0.005 nm, resolution 0.05 nm, acquisition time 10 s.